\documentclass[english, prd, twocolumn, aps, showpacs, amsmath, amssymb, floatfix,nofootinbib]{revtex4-1}

\pdfoutput=1
\usepackage[pdftex]{graphicx}
\usepackage[pdftex]{color}

\usepackage{verbatim}
\usepackage[us]{datetime}

\usepackage[colorlinks,bookmarks]{hyperref}
\definecolor{linkblue}{rgb}{0,0,0.8}
\definecolor{linkgreen}{rgb}{0,0.5,0}
\hypersetup{linkcolor=linkblue, citecolor=linkgreen, urlcolor=linkblue}

\def\mf{\mathbf}

\begin{document}

\title{Weak Lensing Observables in the Halo Model}

\author{Kimmo Kainulainen} \email{kimmo.kainulainen@jyu.fi}
\author{Valerio Marra} \email{valerio.marra@me.com}

\affiliation{Department of Physics, University of Jyv\"{a}skyl\"{a}, PL 35 (YFL), FIN-40014 Jyv\"{a}skyl\"{a}, Finland}
\affiliation{Helsinki Institute of Physics, University of Helsinki, PL 64, FIN-00014 Helsinki, Finland}

\begin{abstract}

The halo model (HM) describes the inhomogeneous universe as a collection of halos. The full nonlinear power spectrum of the universe is well approximated by the HM, whose prediction can be easily computed without lengthy numerical simulations. This makes the HM a useful tool in cosmology. Here we explore the lensing properties of the HM by use of the stochastic gravitational lensing (sGL) method. We obtain for the case of point sources exact and simple integral expressions for the expected value and variance of the lensing convergence, which encode detailed information about the internal halo properties. In particular a wide array of observational biases can be easily incorporated and the dependence of lensing on cosmology is properly taken into account. This simple setup should be useful for a quick calculation of the power spectrum and the related lensing observables, which can play an important role in the extraction of cosmological parameters from, e.g., SNe observations. Finally, we discuss the probability distribution function of the HM which encodes more information than the first two moments and can more strongly constrain the large-scale structures of the universe. To check the accuracy of our modelling we compare our predictions to the results from the Millennium Simulation.

\end{abstract}

\keywords{Gravitational Lenses, Inhomogeneous Universe Models, Observational Cosmology}
\pacs{98.62.Sb, 98.65.Dx, 98.80.Es}


\maketitle

\section{Introduction}

Gravitational lensing by large-scale inhomogeneities affects the light from distant objects in a way which essentially depends on size and composition of the structures through which the photons pass on their way from source to observer.
The fundamental quantity describing this statistical (de)magnification is the lensing probability distribution function (PDF), which has to be understood well if one is to use cosmological observations to accurately map the expansion history and determine the precise composition of the universe (see, for example, \cite{Holz:1998cp,Kim:2003mq, Dodelson:2005zt, Amendola:2010ub, Amanullah:2010vv}).
It is not currently possible to extract the lensing PDF from the observational data and we have to resort to theoretical models.
Two possible alternatives have been followed in the literature.
A first approach (e.g.~\cite{Valageas:1999ir, Munshi:1999qw, Wang:2002qc, Das:2005yb}) relates a ``universal'' form of the lensing PDF to the variance of the convergence, which in turn is fixed by the amplitude of the power spectrum, $\sigma_{8}$. Moreover the coefficients of the proposed PDF may be trained on some specific N-body simulations. A second approach (e.g.~\cite{Holz:1997ic, Bergstrom:1999xh, Holz:2004xx}) is to build {\em ab-initio} a model for the inhomogeneous universe and directly compute the relative lensing PDF, usually through time-consuming ray-tracing techniques. The flexibility of this method is therefore penalized by the increased computational time.

In Refs.~\cite{Kainulainen:2009dw,Kainulainen:2010at} we introduced the stochastic gravitational lensing (sGL) method which combines the flexibility in modeling with a fast performance in obtaining the lensing PDF.
The sGL method is based on the weak lensing approximation and generating stochastic configurations of inhomogeneities along the line of sight.
A numerical code based on sGL, as the publicly available \texttt{turboGL} package \cite{turboGL}, can compute the lensing probability distribution function for a given inhomogeneous model in a few seconds.
The speed gain is actually a sine-qua-non for likelihood approaches, in which one needs to scan many thousands different models (see, e.g.,~\cite{Amendola:2010ub}).
This makes sGL a useful tool to study how lensing depends on cosmological parameters and how it impacts the observations. The method can also be used to simulate the effect of a wide array of systematic biases on the observable PDF.

Here we develop a self-consistent setup to easily calculate matter power spectrum and lensing properties of the desired model universe.
The basic idea is to apply the sGL method to the so-called ``halo model'' (HM) (see, for example, \cite{Neyman:1952, Peebles:1974, Scherrer:1991kk, Seljak:2000gq, Ma:2000ik, Peacock:2000qk, Scoccimarro:2000gm,Cooray:2002dia}), where the inhomogeneous universe is approximated as a collection of different types of halos whose positions satisfy the linear power spectrum. We will give explicit integral expressions for mean and variance of the lensing convergence, which will be directly related to the relative matter power spectrum.
These results will be valid for point sources, while extended sources will be treated in  forthcoming work. Our discussion applies, for example, to the narrow light bundles emitted by distant Supernovae (SNe).
We expect indeed that our formulas will be useful in the analysis of SNe observations as the convergence variance can be straightforwardly included in the standard $\chi^{2}$ analyses.
We also discuss the PDF of the HM which encodes more information than the mean and variance and can more strongly constrain the large-scale structures of the universe.

This paper is organized as follows. In Section~\ref{hm} we introduce the HM and the basic formalism to calculate its power spectrum, while in Section~\ref{lenss} we will discuss its lensing properties.
To check the accuracy of our modelling we compare in Section~\ref{cms} our predictions to the results from the Millennium Simulation (MS)~\cite{Springel:2005nw}.
Finally, we give our conclusions in Section~\ref{concl}.

\section{The Halo Model} \label{hm}

The halo model assumes that on small scales (large wavenumbers $k$) the statistics of matter correlations are dominated by the internal halo density profiles, while on large scales the halos are assumed to cluster according to linear theory. In other words, the nonlinear evolution is assumed to produce only concentrated halos. The model does not include intermediate low density structures such as filaments and walls. The two components are then combined together. We do it here by simple addition so that the total power spectrum is
\begin{equation} \label{pkt}
P(k,z) = P_{L}(k,z) + P_{H}(k,z) \,.
\end{equation}
The first term on the right-hand side is also called 2-halo component and the second term 1-halo component. The usefulness of the halo model stems from the fact that both terms in Eq.~(\ref{pkt}) can be computed without having to resort to numerical simulations.
We would like to stress that more sophisticated and accurate versions of the HM are available in the literature \cite{Smith:2002dz}. Here we use the simplest version of the HM as we merely wish to illustrate its good agreement with the full power spectrum.

\subsection*{Linear Power Spectrum}

The linear part of the power spectrum is as usual:
\begin{equation} \label{pkl}
 {k^{3} \over 2 \pi^{2}} P_{L}(k, z)  = \delta_{H0}^{2}  \left( {c k \over a_{0}H_{0}} \right )^{3+n_{s}} T^{2}(k/a_{0}) D^{2}(z) \,,
\end{equation}
where $H_{0}$ and $a_{0}$ are the present-day Hubble parameter and scale factor, $n_{s}$ is the spectral index and $\delta_{H0}$ is the amplitude of perturbations on the horizon scale today, which we fix by setting $\sigma_{8}$.
For the transfer function $T(k)$ one can use the fits provided, e.g., by Ref.~\cite{Eisenstein:1997ik}.
Fit functions for the growth function $D(z)$ could be found, e.g., in Refs.~\cite{Percival:2005vm,Basilakos:2003bi}, but it can also be easily obtained numerically~\cite{Kainulainen:2010at}.

\subsection*{Halo Power Spectrum}

The nonlinear or halo contribution has been obtained in, e.g., \cite{Scherrer:1991kk}. We will briefly summarize the important steps here. First we introduce the halo mass function $f(M,z)$, which gives the fraction of the total mass in halos of mass $M$ at the redshift $z$. The function $f(M,z)$ is related to the number density $n(M,z)$ by:
\begin{equation}
{\rm d}n(M,z) \equiv n(M,z) {\rm d}M 
                = {\rho_{MC} \over M} \, f(M,z) {\rm d}M \, ,
\label{eq:halomassf}
\end{equation}
where the density $\rho_{MC} \equiv a_0^3 \, \rho_{M0}$ is the constant matter density in a co-moving volume. The halo function is by definition normalized to unity: $\int f(M,z) {\rm d}M = 1$ and we defined ${\rm d}n$ as the number density of halos in the mass range ${\rm d}M$. The power spectrum due to randomly located halos is then~\cite{Peacock:2000qk}
\begin{equation} \label{pkh}
P_{H}(k,z)=  \int_{0}^{\infty} {\rm d}n(M,z)  
\left({M \, W_k(M,z) \over \rho_{MC}} \right)^{2}   \,,
\end{equation}
where $W_k$ is the Fourier transform of the halo density profile:
\begin{equation}
W_k(M,z)=  \frac{1}{M} \int_{0}^{R} \rho(r,M,z)  {\sin k r \over k r}  4\pi \, r^2 \, {\rm d}r \,,
\end{equation}
and $R$ is the halo radius. For the halo profile $\rho(r,M,z)$, we shall use the Navarro-Frenk-White (NFW) profile~\cite{Navarro:1995iw}. We will now apply this setup to the weak lensing.

\section{Lensing} \label{lenss}

The lens convergence $\kappa$ in the weak-lensing approximation is given by the following integral evaluated along the unperturbed light path~\cite{Bartelmann:1999yn}:
\begin{equation} \label{eq:kappa}
    \kappa(z_{s})=\int_{0}^{r_{s}}dr \, \rho_{MC} \, G(r,r_{s})\,\delta_{M}(r,t(r))
\end{equation}
where we defined the auxiliary function
\begin{equation} \label{opw}
G(r,r_{s})=  \frac{4\pi G}{c^2  a}  \, \frac{f_{k}(r)f_{k}(r_{s}-r)}{f_{k}(r_{s})} \,,
\end{equation}
which gives the optical weight of an inhomogeneity at the comoving radius $r$.
The functions $a(t)$ and $t(r)$ are the scale factor and geodesic time for the background FLRW model, and $r_{s}=r(z_{s})$ is the comoving position of the source at redshift $z_{s}$. 
Also, $f_{k}(r)=\sin(r\sqrt{k})/\sqrt{k},\, r,\,\sinh(r\sqrt{-k})/\sqrt{-k}$ depending on the curvature $k>,=,<0$, respectively.
Neglecting the second-order contribution of the shear, the shift in the distance modulus caused by lensing is expressed solely in terms of the convergence:
\begin{equation} \label{eq:dm}
\Delta m(z) \simeq  5 \log_{10}(1-\kappa(z)) \,.
\end{equation}
Eqs.~(\ref{eq:kappa}) and (\ref{eq:dm}) show that for a lower-than-FLRW column density the light is demagnified (e.g., empty beam $\delta_{M}=-1$), while in the opposite case it is magnified.
In Eq.~(\ref{eq:kappa}) the quantity $\delta_{M}(r,t)$ is the local matter density contrast and it is clear that an accurate statistical modelling of the magnification PDF requires a detailed description of the inhomogeneous mass distribution. In this work we will model $\delta_{M}$ according to the halo model. See \cite{Kainulainen:2010at} for a more refined modelling which also includes filamentary structures confining the halos.
The following results will be valid for point sources (smoothing angle $\theta =0$).
We will discuss extended sources in a forthcoming paper.

Because in Eq.~(\ref{eq:kappa}) the contributions to the total convergence are combined additively, it is useful to decompose the density field into a sum of Fourier modes:
\begin{equation} \label{dek}
\delta_{M}(\mf{r},t) = \int_0^{\infty} {d^{3} \mf{k} \over (2 \pi)^{3}} \, e^{i \mf{k} \cdot \mf{r}} \, \delta_{M}(\mf{k},t) \,.
\end{equation}
Eq.~(\ref{dek}) can be used to separate the contributions to the convergence due to small-scale inhomogeneities ($k \gtrsim k_{\rm cut}$) from the ones due to large-scale inhomogeneities ($k \lesssim k_{\rm cut}$) so that, similarly to Eq.~(\ref{pkt}), it is:
\begin{equation} \label{kappat}
\kappa(z) = \kappa_{L}(z) + \kappa_{H}(z) \,.
\end{equation}
The quantity $k_{\rm cut}$ separates the modes relevant to $\kappa_{L}$ from the ones relevant to $\kappa_{H}$ and can be defined as the scale at which the two power-spectrum components have the same value, $P_{L}(k_{\rm cut},z) = P_{H}(k_{\rm cut},z)$.

We are in particular interested in the expected value and variance of the lensing convergence. From Eq.~(\ref{kappat}) it follows that:
\begin{eqnarray}
\langle \kappa \rangle &=& \langle \kappa_{L} \rangle +\langle \kappa_{H} \rangle \,, \label{meankt} \\
\sigma^{2}_{\kappa} &=& \sigma^{2}_{\kappa_{L}} + \sigma^{2}_{\kappa_{H}}  \label{varkt} \,.
\end{eqnarray}
We will now give directly computable expressions for Eqs.~(\ref{meankt}-\ref{varkt}).

\subsection*{Linear Contribution}

Because of photon conservation, if no lines of sight are obscured, the expected value of the weak lensing convergence is zero. This can be seen in Eq.~(\ref{eq:kappa}) from the fact that $\langle \delta_{M}(r,t) \rangle = 0$. Mechanisms that are able to obscure lines of sight involve small scales and so we have
\begin{equation}
\langle \kappa_{L} \rangle = 0 \,.
\end{equation}
The linear power spectrum, however, contributes to the variance of the convergence according to \cite{Bernardeau:1996un}:
\begin{equation} \label{varkappal}
\sigma^{2}_{\kappa_{L}}=   \int_{0}^{r_{s}} dr \rho_{MC}^{2}G^{2}(r,r_{s}) \int_0^{k_{\rm cut}} {k \, dk \over 2 \pi}  P_{L}(k,z(r)) \,,
\end{equation}
where the scale $k_{\rm cut}$ can depend on the redshift $z$.

\subsection*{Halo Contribution}

The sGL method for computing the lensing convergence is based on generating stochastic configurations of halos and filaments along the line of sight and computing the associated integral in Eq.~(\ref{eq:kappa}) by binning into a number of independent lens planes. Because the halos are randomly placed their occupation numbers in parameter-space volume cells follow Poisson statistics. This allows rewriting Eq.~(\ref{eq:kappa}) as a sum over these cells characterized by the corresponding Poisson occupation numbers.
The various individual contributions to the convergence are then additively combined. By generating many halo configurations one can easily sample the convergence PDF. Besides the approximations relative to Eqs.~(\ref{eq:kappa}-\ref{eq:dm}), correlations in the halo positions are neglected here, similarly to Eq.~(\ref{pkh}). In the full sGL one can model also filamentary structures confining the halos, thus accounting for some of the correlations among the halo positions. A detailed explanation of the sGL method can be found in \cite{Kainulainen:2010at,Kainulainen:2009dw} and a publicly-available numerical implementation, the \texttt{turboGL} package, in \cite{turboGL}. 

In the present case the filamentary structures are turned off and the inhomogeneous matter distribution introduced in \cite{Kainulainen:2010at} exactly corresponds to the halo model. The sGL method then predicts the following direct and exact results for mean and variance of the convergence:
\begin{eqnarray} \label{averagekappax}
 \langle \kappa_{H} \rangle &=&    \int_{0}^{r_{s}} {\rm d}r \, G(r,r_{s}) \int_{0}^{\infty} {\rm d}n(M,z(r)) \times \\
& \times& \int_{0}^{R(M,z(r))}{\rm d}A(b) \,  (P_{\rm sur}  -1) \,  \Sigma(b,M,z(r))  \,,  \nonumber
\end{eqnarray}
and
\begin{eqnarray} \label{varkappax}
 \sigma^{2}_{\kappa_{H}}&=&    \int_{0}^{r_{s}} {\rm d}r \, G^{2}(r,r_{s}) \int_{0}^{\infty} {\rm d}n (M,z(r)) \times \\
& \times& \int_{0}^{R(M,z(r))}{\rm d}A(b)  \, P_{\rm sur}   \, \Sigma^{2}(b,M,z(r))  \,, \nonumber
\end{eqnarray}
where the integral limits for the last two integrals are implicitly defined, ${\rm d}A(b) \equiv 2 \pi b \, {\rm d}b$ and $\Sigma$ is the halo surface density:
\begin{equation}
\Sigma(b,M,z) = a^{3} \int_{b}^{R} \frac{2 \,   r \, {\rm d}r}{\sqrt{r^2-b^2}} \, \rho(r,M,z) \,.
\end{equation}
Eqs.~(\ref{averagekappax}-\ref{varkappax}) can be directly integrated without having to consider the full formalism of the sGL method.
We also would like to point out that, in contrast to, e.g., \cite{Cooray:2000zy}, our expressions use the halo profiles in real space and not in Fourier space, thus including higher order correlation terms beyond the power spectrum. As a direct consequence the survival probability $P_{\rm sur}$ can be included in the way the sGL method predicts.

The quantity $P_{\rm sur}=P_{\rm sur}(b,M,z,z_{s})$ is a generic function that describes the probability that a light ray, which encounters a halo of mass $M$ at the redshift $z$ and impact parameter $b$ for a source at redshift $z_{s}$, does not fall below detection threshold. $P_{\rm sur}$ can be used to simulate the effect of a very wide array of systematic biases, such as any sources leading to obscuration of the light beam, either alone or in combination with restrictions arising from imperfect search efficiencies and strategies. For example, selection by extinction effects or by outlier rejection mainly relates to high-magnification events which are clearly correlated with having high intervening mass concentrations (halos with large $M$ and/or small $b$) along the light geodesic. Similarly, short duration events might be missed by search telescopes or be rejected from the data due to poor quality of the light curve, for example in cases where a supernova is not separable from the image of a bright foreground galaxy. Probability of such events would be correlated with the brightness of the source, with the density and redshift of the intervening matter and of course with the search efficiencies. All these effects could be modelled by  $P_{\rm sur}$, carefully adjusted by the use of the astrophysical input.
As a simple illustration, in the next Section we will model the survival probability by a step function in the impact parameter, such that the halo is opaque for radiuses smaller than the NFW scale radius $R_{s}= R/c$:
\begin{equation} \label{toysel}
P_{\rm sur}(b,M,z)= \left\{
  \begin{array}{ll}
    0 & \; b/R< c(M,z)^{-1} \\
    1 & \; {\rm otherwise} 
  \end{array}\right.\; ,
\end{equation}
where $c(M,z)$ is the mass- and redshift-dependent NFW concentration parameter.

Eqs.~(\ref{averagekappax}-\ref{varkappax}) allow to draw some general considerations. First, if $P_{\rm sur}=1$, the expected value of the convergence is correctly zero as demanded by photon flux conservation,\footnote{This should be approximately true~\cite{Weinberg:1976jq} if strong lensing events leading to secondary images and caustics~\cite{Ellis:1998ha} do not play a significant role as far as the full-sky average is concerned~\cite{Holz:1997ic,Hilbert:2007ny,Hilbert:2007jd}.} showing the ``benevolent'' nature of weak lensing corrections for unbiased observations.
If, however, the survival probability is not trivial, selection biases persist even in very large datasets so that the average convergence approaches a nonvanishing value.
Second, Eq.~(\ref{varkappax}) is a product of positive quantities and so {\em a nontrivial survival probability  always reduces the observed variance.}  Thus, neglecting existing systematic biases  could lead to an underestimation of the observationally inferred variance, and hence to too strong constraints on additional variance caused by inhomogeneities.

Finally, Eqs.~(\ref{averagekappax}-\ref{varkappax}) are integrated (see second d$n$-integral) over all halo masses $0 < M < \infty$.
It is, however, straightforward to generalize them by changing the integration limits in order to have expected value and variance relative to a specific halo mass bin $\Delta M$.
This may be useful if one wants to connect the lensing given by halos within a mass bin (see Fig.~\ref{vpb}) to the corresponding amount of correlation.

\section{Comparison with the Millennium Simulation} \label{cms}

\begin{figure}
\begin{center}
\includegraphics[width=8.5 cm]{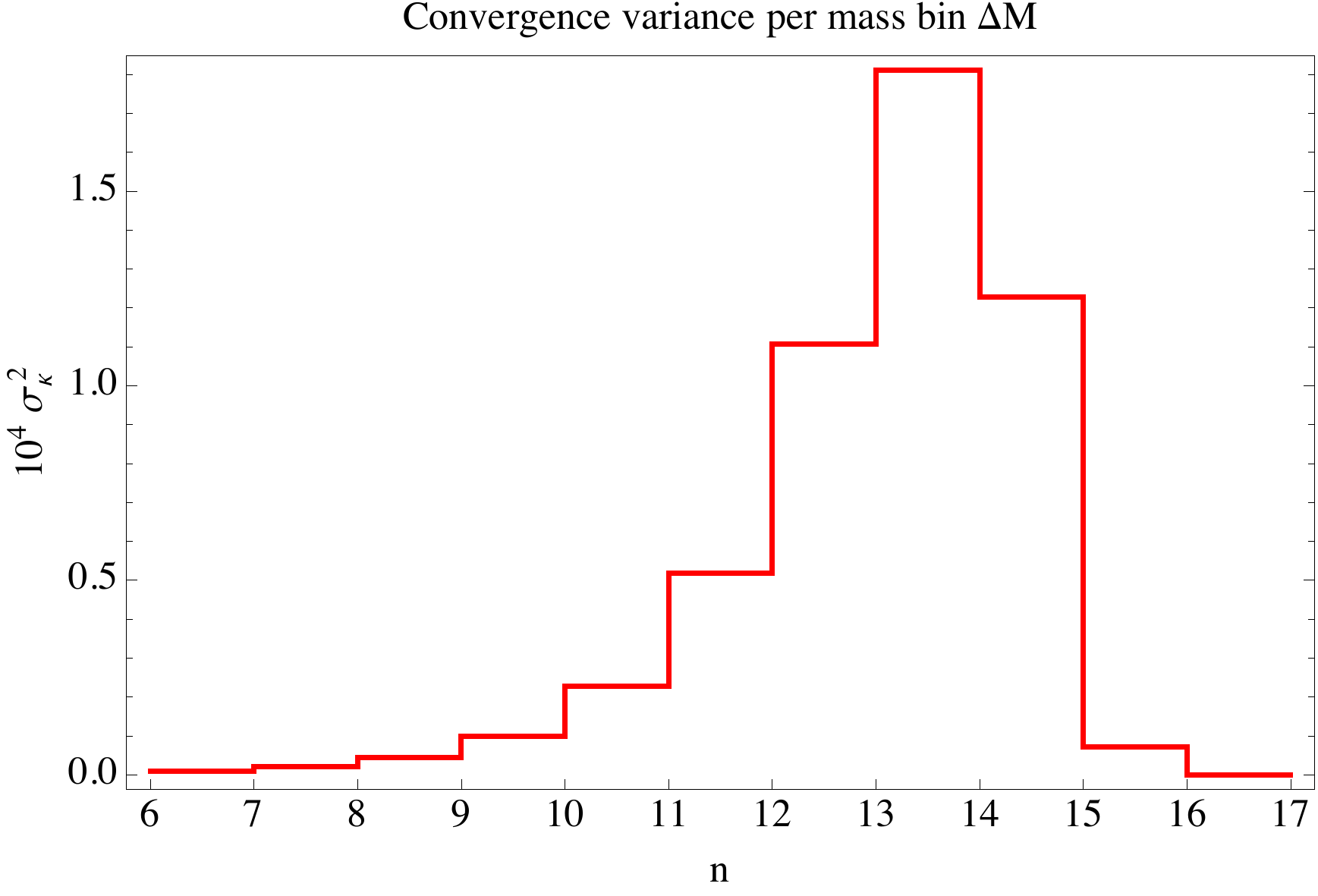}
\caption{
Shown is the variance per mass bin for a source at redshift $z=1$.
The variance is computed using Eq.~(\ref{varkappax}) restricted to the mass bin $\Delta M= ]10^{n},\, 10^{n+1}] \, h^{-1} M_{\odot}$.
In the plot $n$ labels the halo mass according to $M_{n}=10^n h^{-1} M_{\odot}$.
For halos of mass $M \lesssim 10^{9} h^{-1} M_{\odot}$ the contribution to the total variance is negligible. See Section \ref{cms} for more details.
}
\label{vpb}
\end{center}
\end{figure}

We will now compare the halo model power spectrum and lensing properties to those from the Millennium Simulation. Accordingly, we will fix the cosmological parameters to
$h=0.73$, $\Omega_{M}=0.25$,  $\Omega_{B}=0.045$, $\Omega_{\Lambda}=0.75$, $w=-1$, $\sigma_{8}=0.9$ and $n_{s}=1$; see \cite{Springel:2005nw} for more details.
Let us point out that in \cite{Kainulainen:2010at} we have already accurately reproduced the MS lensing PDF with an sGL modelling which included not only the halos, but also the filamentary structures. Our goal here is different; we are deliberately using a simpler modelling of the inhomogeneities in order to study the validity of the halo model for lensing analyses.

Because we are assuming that all (virialized) matter is concentrated in halos, we need to use a halo mass function whose integral is normalizable to unity. We therefore adopt the halo function provided by Sheth \& Tormen in Ref.~\cite{Sheth:1999mn}, which should be approximately valid for the full mass range of the integrals in Eqs.~(\ref{averagekappax}-\ref{varkappax}).
Halo functions obtained through numerical simulations, albeit possibly more precise, are valid only above a mass cutoff imposed by the numerical resolution of the simulation itself~\cite{Jenkins:2000bv}.
We point out, however, that halos with mass smaller than $M_{\rm cut} \sim 10^{9} h^{-1} M_{\odot}$ act effectively as a mean field in weak lensing, and so mass functions from numerical simulations which are valid down to $M_{\rm cut}$ may be used within the present setup by introducing explicitly $M_{\rm cut}$ in Eqs.~(\ref{averagekappax}-\ref{varkappax}).
This fact can be appreciated quantitatively by computing the convergence variance (\ref{varkappax}) per mass bin $\Delta M$ as shown in Fig.~\ref{vpb}. Clearly for $M \lesssim M_{\rm cut}$ the contribution to the total variance is negligible which means that the corresponding halos behave effectively as a fine homogeneous dust as far as weak lensing is concerned~\cite{Kainulainen:2010at}.

\begin{figure}
\begin{center}
\includegraphics[width=8.5 cm]{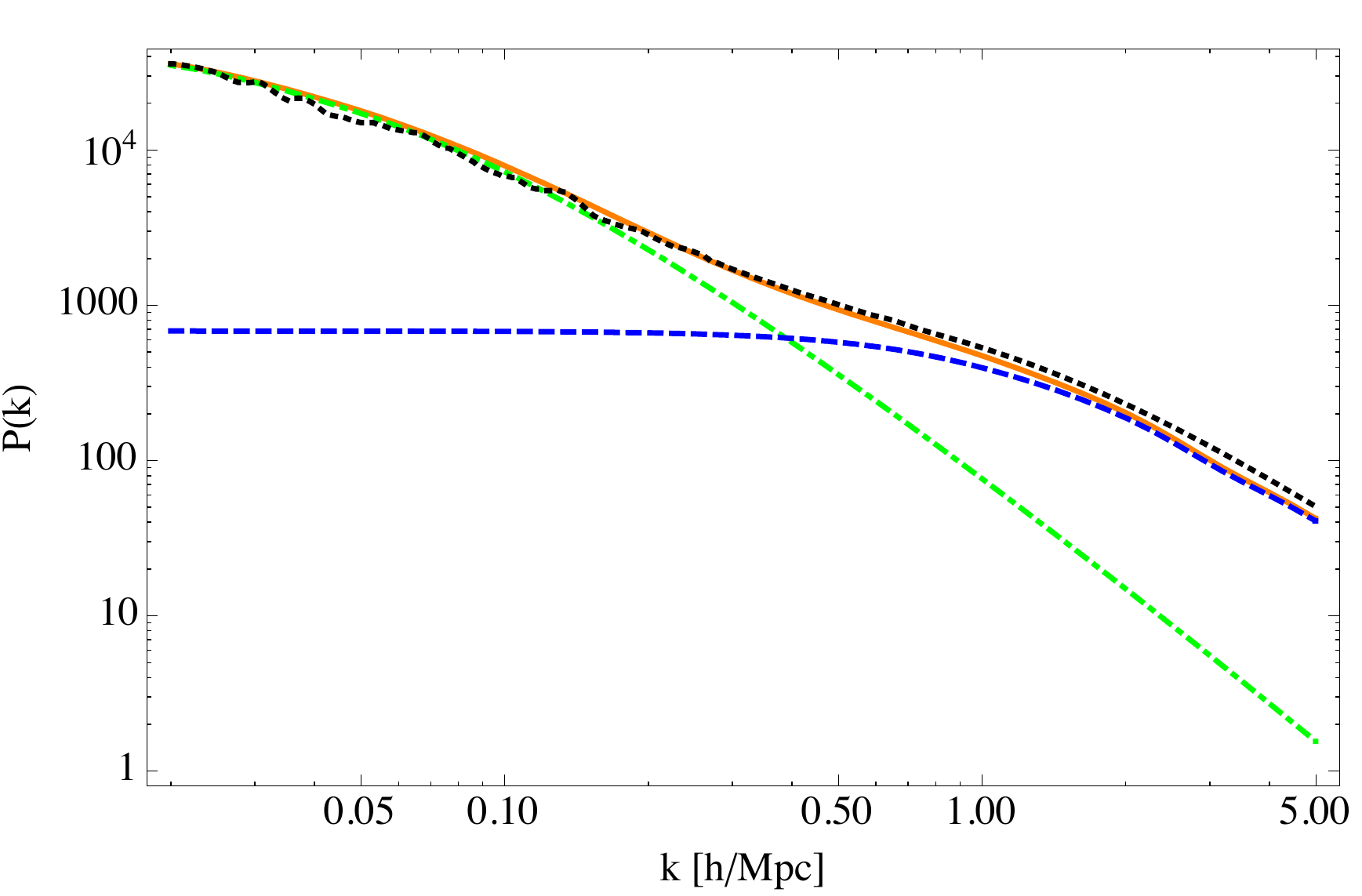}
\caption{Shown is the present-day power spectrum of the halo model (orange solid line and Eq.~(\ref{pkt})) for the $\Lambda$CDM universe of the Millennium Simulation, together with the halo (blue dashed line and Eq.~(\ref{pkh})) and linear (green dot-dashed line and Eq.~(\ref{pkl})) components.
Note how the halo component gives a constant contribution at large scales (shot noise).
Also shown for comparison is the power spectrum (black dotted line) relative to the Millennium Simulation \cite{Springel:2005nw}.}
\label{Pk}
\end{center}
\end{figure}

In Fig.~\ref{Pk} we have plotted the power spectrum of Eq.~(\ref{pkt}), together with the halo and linear components. Also plotted is the power spectrum of the Millennium Simulation. The agreement is rather good, although the halo model power spectrum is slightly underpowered on small scales.
A better power spectrum could be obtained with a more sophisticated HM, but this would not be relevant for the main goal of this paper, which is to propose using Eqs.~(\ref{averagekappax}-\ref{varkappax}) for lensing.
The oscillations at large scales are absent in the HM spectrum because we used a  transfer-function fit that  reproduces the baryon-induced suppression on the intermediate scales but ignores the acoustic oscillations, which are again not relevant for us here \cite{Eisenstein:1997ik}.
For the NFW concentration parameters we used the fit\footnote{We increased the fit by a factor of $4.67/3.93$ (see \cite{Duffy:2008pz}) in order to match the results of \cite{Neto:2007vq} which are relative to the MS cosmology.} of Ref.~\cite{Duffy:2008pz}.
Note also that, as said before, it is straightforward to calculate the correlation per mass bin by binning Eq.~(\ref{pkh}) in the same way Eq.~(\ref{varkappax}) has been binned in Fig.~\ref{vpb}.

\begin{figure}
\begin{center}
\includegraphics[width=8.5 cm]{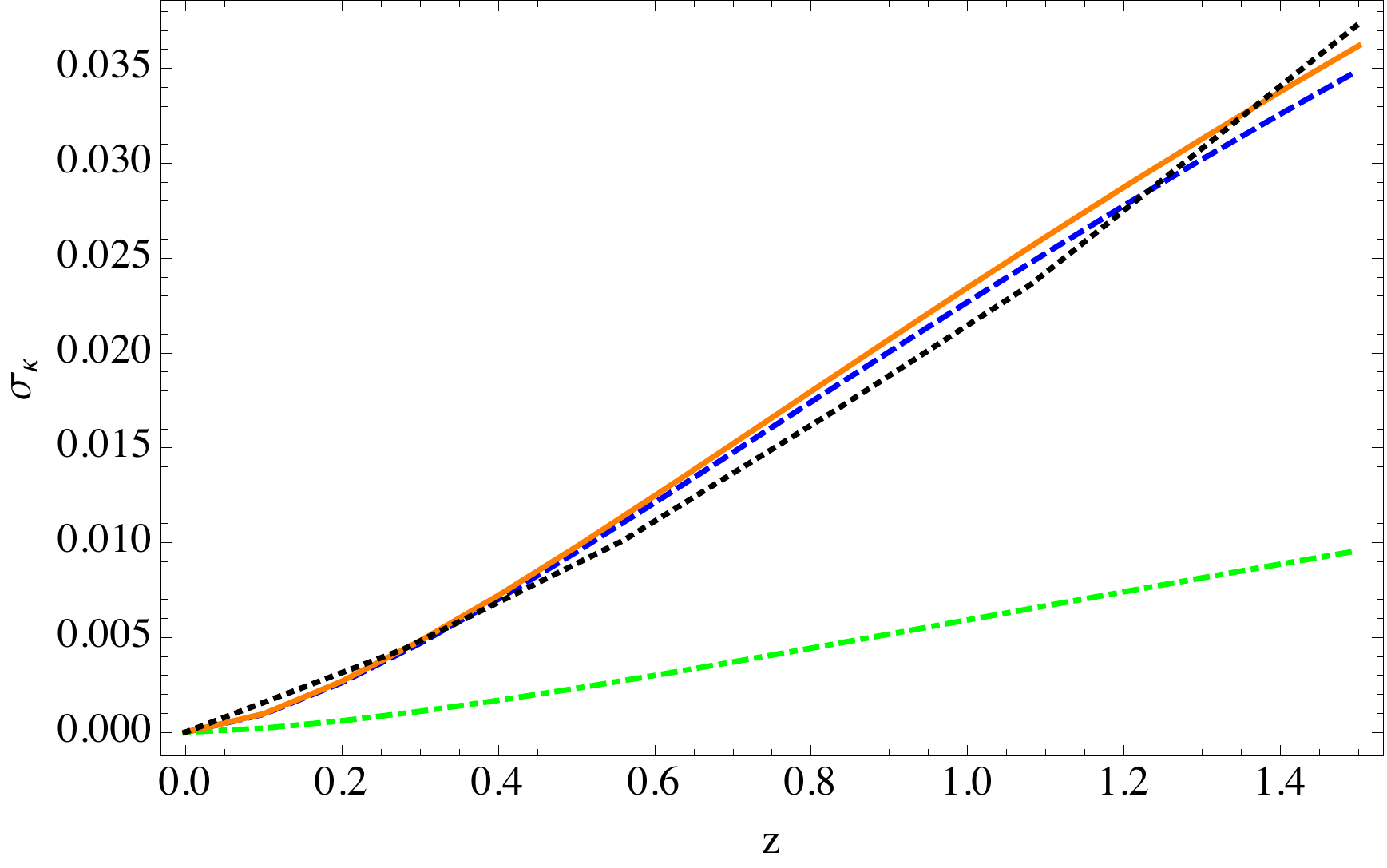}
\caption{Shown is the redshift dependence of the dispersions due to the linear (green dot-dashed line and Eq.~(\ref{varkappal})) and halo (blue dashed line and Eq.~(\ref{varkappax})) components, for the $\Lambda$CDM model of the Millennium Simulation. Also plotted is the sum (orange solid line) of the variances and the dispersion (black dotted line) relative to the Millennium Simulation~\cite{Hilbert:2007ny, Hilbert:2007jd}. The labeling is as in Fig.~\ref{Pk}.}
\label{sigmas}
\end{center}
\end{figure}

In Fig.~\ref{sigmas} we have plotted the redshift dependence of the convergence dispersions due to the linear and halo components, computed with Eq.~(\ref{varkappal}) and Eq.~(\ref{varkappax}), and also the sum of the variances which shows that the linear contribution is practically negligible. 
We wish to stress that here we are studying the case of point sources, i.e., we are adopting a smoothing angle $\theta =0$. For extended sources and/or finite smoothing angles the results of Fig.~\ref{sigmas} change, and for smoothing angles larger than some value $\bar \theta$ the linear contribution dominates over the one due to the halos. At present we cannot treat nonzero smoothing angles within the sGL model, but we can nonetheless estimate the order of magnitude of $\bar \theta$.
Assuming for simplicity a flat universe, the optical weight function $G$ of Eq.~(\ref{opw}) peaks at half the comoving distance to the source and  vanishes at the observer and source locations. The physics at $r_{s}/2$ is therefore the most relevant for lensing, and a smoothing angle $\theta$ can  be converted into a smoothing scale by means of $\lambda=\theta \, d_{A}(r_{s}/2)$, where $d_{A}$ is the angular diameter distance. Let us now take, for example, a source at $z=1$ and a smoothing scale of $\bar \lambda \sim 1$ Mpc, which should be enough to smooth out the inhomogeneous halo profiles. For the MS cosmology the corresponding smoothing angle is $\bar \theta \sim 3^{\prime}$.
Also plotted in Fig.~\ref{sigmas} is the dispersion relative\footnote{In order not to exceed the validity region of the weak-lensing approximation, the dispersion relative to the MS has been calculated in the weak-lensing regime $\kappa \ll 1$ by applying a cut at large magnifications. More precisely, to ensure that the PDF is basically unaltered, the cut has been determined by asking that the high-magnification tail has the total probability of $10^{-3}$. This resulted in $\kappa_{\rm cut}=0.063, 0.127, 0.195, 0.255, 0.365$ for the PDFs relative to $z=0.28, 0.56, 0.83, 1.08, 1.50$, respectively.} to the Millennium Simulation~\cite{Hilbert:2007ny, Hilbert:2007jd}, which shows a good agreement with the theoretical result.
We wish to stress that it is very easy to evaluate the integrals of Eqs.~(\ref{averagekappax}-\ref{varkappax}) numerically, and their predictions can be straightforwardly implemented in $\chi^{2}$ analyses based on Gaussian likelihoods. However, it should be remembered that Gaussian analysis is justified if the lensing PDF is not strongly skewed. When this is not the case the full PDF  has to be used for the likelihood analysis, as was done, e.g., in \cite{Amendola:2010ub}.

\begin{figure}
\begin{center}
\includegraphics[width=8.5 cm]{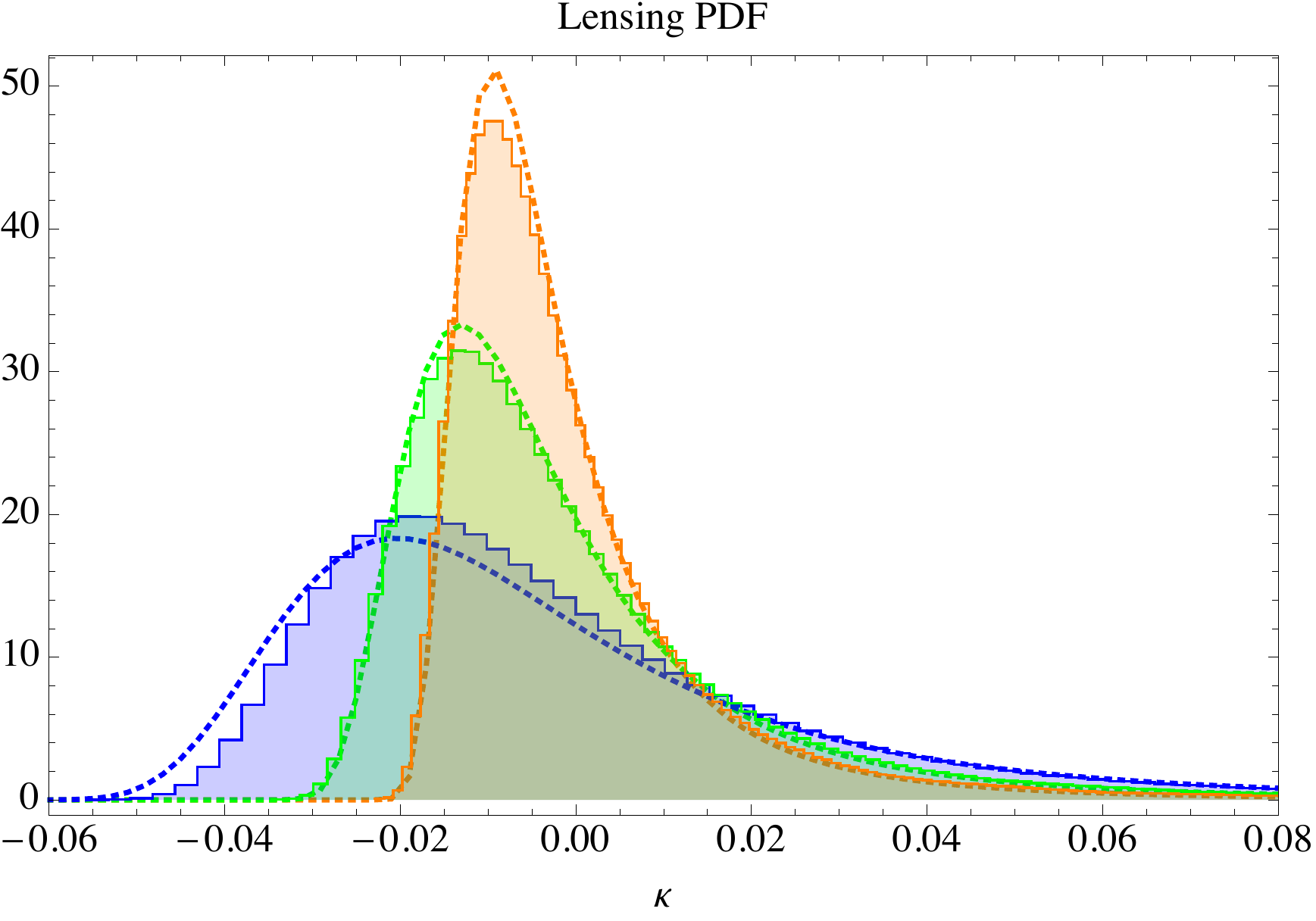}
\caption{Shown are the lensing PDFs for a source at (from taller to shorter graph) $z=0.83,\, 1.08$ and $1.50$ for the $\Lambda$CDM model of the Millennium Simulation.
The dotted lines are the  lensing PDFs generated by shooting rays through the MS \cite{Hilbert:2007ny, Hilbert:2007jd}, while the histograms are the corresponding lensing PDFs obtained with the sGL method~\cite{Kainulainen:2010at,Kainulainen:2009dw,turboGL}.}
\label{PDF}
\end{center}
\end{figure}

In Fig.~\ref{PDF} we have plotted the lensing PDFs obtained with the sGL method (histograms).
Only the (1-)halo contribution is included as this is the direct output of the sGL method. Indeed at present we cannot properly include the linear contribution which, in any case, should be subdominant as shown in Fig.~\ref{sigmas}.
For $z=0.83$ and $1.08$ the agreement with the MS results (dotted lines) from \cite{Hilbert:2007ny, Hilbert:2007jd} is remarkable, especially  considered the simplicity of the setup.
For $z=1.50$ the halo model predicts a less skewed PDF.
This could mean that the model we are using to describe the halos (Sheth \& Tormen mass function and NFW concentration parameters from~\cite{Duffy:2008pz}) is less accurate at high redshifts. For example, a better agreement with \cite{Hilbert:2007ny, Hilbert:2007jd} is found if the concentration parameters are increased by $40\%$, and similar results may be found by slightly changing the shape of the mass function.
 Alternatively the extra skewness could be caused by unvirialized 
 low-density large-scale structures like filaments and walls. These 
 could be particularly important at high redshifts where lesser amounts
 of virialized structures, which are the only ones accounted for by the halo model, are present. 
Indeed, a very good agreement with MS was  found in \cite{Kainulainen:2010at} where the halo model was extended to include filamentary structures.
Note, finally, that the differences in shape between the two PDFs at $z=1.50$ do not seem to strongly impact the dispersions plotted in Fig.~\ref{sigmas}.

\subsection*{Selection effects}

\begin{figure}
\begin{center}
\includegraphics[width=8.5 cm]{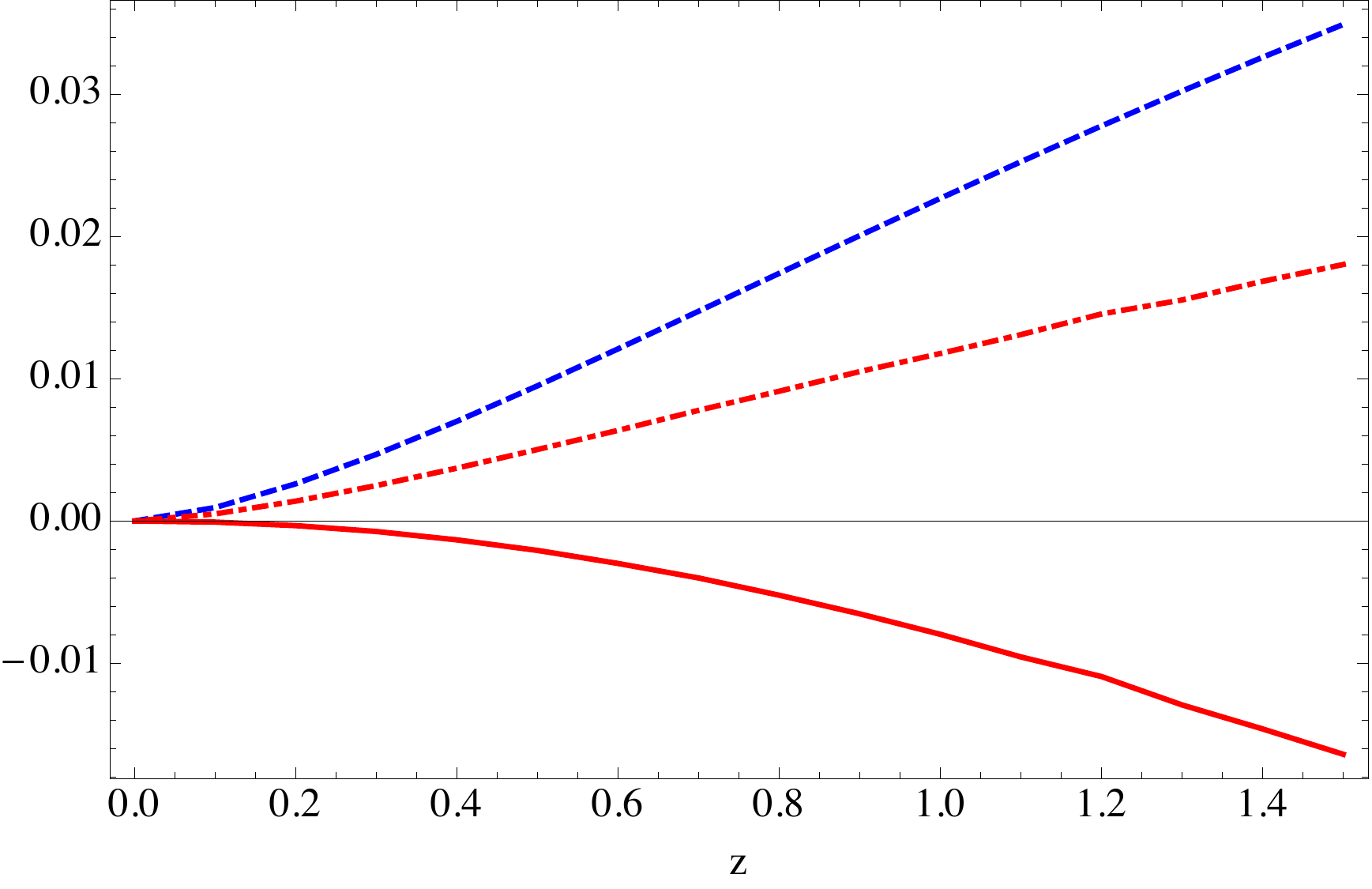}
\caption{Shown is the redshift dependence of the convergence dispersion with (red dot-dashed line) and without (blue dashed line) the selection effects of Eq.~(\ref{toysel}), for the $\Lambda$CDM model of the Millennium Simulation.
Also plotted is the mean convergence (red solid line) with selection effects (without selection effects it is  zero).}
\label{smsur}
\end{center}
\end{figure}

It is interesting to discuss the impact on lensing of the toy survival probability of Eq.~(\ref{toysel}).
In Fig.~\ref{smsur} we show the redshift dependence of the convergence dispersion (\ref{varkappax}) with (dot-dashed line) and without (dashed line) the selection effect, which halves the resulting dispersion.
Also plotted is the (negative) mean convergence (\ref{averagekappax}), which shows the possible bias present even in very large datasets.

\begin{figure}
\begin{center}
\includegraphics[width=8.5 cm]{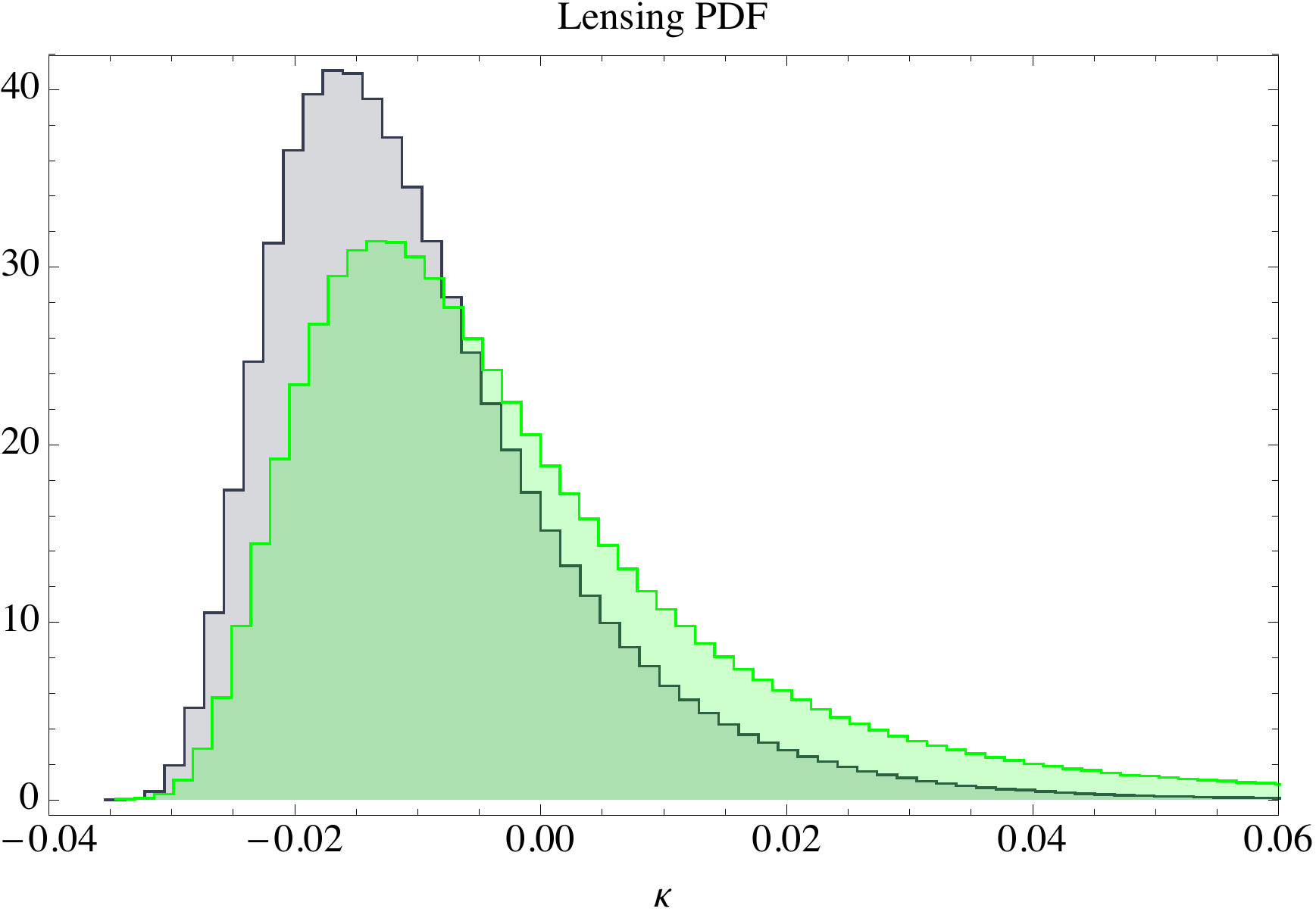}
\caption{Shown is the lensing PDF for a source at $z=1.08$ with (taller histogram) and without (shorter histogram) the selection effects of Eq.~(\ref{toysel}), for the $\Lambda$CDM model of the Millennium Simulation.}
\label{PDFsur}
\end{center}
\end{figure}

In Fig.~\ref{PDFsur} we show the full lensing PDF with selection effects included (taller PDF). The lensing PDF with unitary survival probability is also shown for comparison (shorter PDF).
As expected the selection effects reduce the high magnification tail of the PDF, without sizeably changing the tail at low magnifications.
See \cite{Kainulainen:2010at} for another discussion of the impact of selection effects on lensing.

Finally, we would like to point out that a finite smoothing angle reduces the skewness of the PDF as it reduces the contrast in the matter surface density.
If observational biases are added, however, the overall effect on the PDF is more involved and depends strongly on the kind of selection effect considered.
In the case of obscuration, for example, the bias on the mean convergence should remain basically unchanged as photons relative to unseen lines of sight are lost no matter the smoothing angle adopted. Therefore for increasing $\theta$ the PDF should tend to a gaussian with decreasing dispersion peaked at the mean biased convergence.

\section{Conclusions} \label{concl}

By applying the sGL method to the halo model, we have obtained a simple setup that allows to quickly compute power spectrum and lensing observables for a desired model universe.
We wish to stress that there are no extra free parameters in the model, besides the ones relative to the halo model derived from the literature (i.e., halo mass  and profile fitting functions).
In particular, we have given exact and directly-computable expressions for the expected value (see Eq.~(\ref{averagekappax})) and variance (see Eq.~(\ref{varkappax})) of the convergence.
We stress that Eqs.~(\ref{averagekappax}-\ref{varkappax}) contain more information than analogous expressions based on the power spectrum, like Eq.~(\ref{varkappal}). They are a direct consequence of the sGL method which models the halo profiles in real space, and therefore include information of coherent structures described by higher order correlation terms beyond the power spectrum. In particular, a wide array of systematic biases, which could be relevant for the extraction of cosmological parameters from, e.g., SNe observations and which persist even in very large data sets, can be included in Eqs.~(\ref{averagekappax}-\ref{varkappax}). We have shown a quantitative example of the impact of selection effects on lensing with Figs.~\ref{smsur} and \ref{PDFsur}.
Moreover, these equations have the crucial advantage of including the dependence of lensing on cosmology, which was shown in \cite{Amendola:2010ub} to give sizeable corrections to the confidence level contours from SNe observations. This is to be contrasted with the usual approach (e.g., \cite{Kowalski:2008ez,Amanullah:2010vv}) where lensing effects are included by means of a {\em cosmology-independent} dispersion, which neglects the important dependence on, e.g., $\sigma_{8}$, $\Omega_{M, \Lambda}$ and $h$.

We have then checked the accuracy of this setup against the cosmology of the Millennium Simulation.
We have found that the theoretical predictions of the variance are in good agreement with the results relative to the Millennium Simulation~\cite{Hilbert:2007ny, Hilbert:2007jd}.
This is most important as we wish here to propose to the community the use of Eqs.~(\ref{averagekappax}-\ref{varkappax}) for lensing, which can be easily included in the standard $\chi^{2}$ analyses.
We have also found that the contribution of the linear (2-halo) component to the total variance is practically negligible.
Then we have  shown the lensing PDFs of the halo model as computed using the  sGL method and the \texttt{turboGL} package \cite{turboGL} for the redshifts of $z=0.83,\, 1.08$ and $1.50$.
Given the simplicity of the setup, we found a remarkable agreement with the results from the Millennium Simulation, especially at $z=0.83$ and $1.08$.
One could conclude that, as far as the weak lensing magnification is concerned, the main contribution comes from the smooth halo profiles, with a smaller correction due to correlations among the halo positions and substructures within the halos (even though one should keep in mind the possibility that the MS itself might lack some interesting substructures due to its resolution limit).
At  $z=1.50$ we have found that the halo model predicts a less-skewed PDF.
This could be either due to the fact that the model we are using to describe the halos (Sheth \& Tormen mass function with NFW halos) is less accurate at high redshifts or due to unvirialized low-density large-scale structures like filaments and walls, which could be important at higher redshifts where less virialized structures are present.
One could indeed improve the modelling of the inhomogeneities, for example by explicitly introducing the filamentary structures as in \cite{Kainulainen:2010at}, or by considering substructures within halos and filaments. Indeed, any extra matter structures such as filaments and walls increase the skewness of the PDF, and it would be interesting to see to what extent one can alter the lensing PDF by introducing such structures without conflicting with the observational constraints on the matter power spectrum.
We will develop these thoughts further in a forthcoming paper.

\begin{acknowledgments}

It is a pleasure to thank Stefan Hilbert for providing the results of Ref.~\cite{Hilbert:2007ny, Hilbert:2007jd} and Mike Boylan-Kolchin for providing the power spectrum of the Millennium Simulation.
The authors benefited from discussions with Luca Amendola and Matthias Bartelmann.

\end{acknowledgments}



\end{document}